\documentclass[
onecolumn,
notitlepage,
superscriptaddress,
]{revtex4-2}
\usepackage[T1]{fontenc}

\usepackage[
    a4paper, 
    mag=1000,
    left=3cm, 
    right=3cm,
    top=2cm,
    bottom=2cm,
    headsep=0.7cm,
    footskip=1cm
    ]{geometry}

\usepackage{setspace}
\usepackage{titlesec}

\titleformat{\section}[block]
{\normalfont\large\bfseries\raggedright}
{\thesection.}
{0.6em}
{}

\titleformat{\subsection}[block]
{\normalfont\normalsize\bfseries\raggedright}
{\thesubsection.}
{0.6em}
{}
    
\usepackage[utf8]{inputenc}
\usepackage{amssymb}
\usepackage{mathtools}
\usepackage[T2A]{fontenc}
\usepackage{amsmath}
\usepackage[russian,english]{babel}
\usepackage{graphicx}
\usepackage[table]{xcolor}

\usepackage{amsfonts} 

\usepackage{enumerate}

\usepackage{graphicx}
\usepackage{dcolumn}
\usepackage{bm}
\usepackage{hyperref}
\usepackage{array}
\usepackage{comment}
\usepackage{float}
\usepackage{subcaption}
\usepackage{xcolor}

\begin{document}

\title{Shift-register synchronization of dual-wavelength pulse trains in resonators with quadratic nonlinearity}

\author{Stepan~Bogdanov}
\thanks{These authors contributed equally to this work.}
\affiliation{\mbox{Aston Institute of Photonic Technologies, Aston University, Birmingham, B4 7ET, UK}}

\author{Anastasiya~Bednyakova}
\thanks{These authors contributed equally to this work.}
\affiliation{Novosibirsk State University,
Novosibirsk, 630090, Russia}

\author{Sergei~K.~Turitsyn}
\affiliation{\mbox{Aston Institute of Photonic Technologies, Aston University, Birmingham, B4 7ET, UK}}

\begin{abstract} 
Self-organization of nonlinear waves underpins phenomena ranging from hydrodynamic pattern formation to optical frequency-comb generation. We introduce and numerically demonstrate a shift-register synchronization mechanism for pulse trains in quadratic nonlinear resonators. The synchronized state persists despite substantial group-velocity mismatch through continuous parametric energy exchange, resulting in deterministic discrete pulse shifts after each cavity round trip. Nonlinear parametric interaction modifies the effective walk-off, allowing the pulse trains to satisfy the shift-register synchronization conditions. The proposed mechanism reveals a previously unexplored synchronization regime in quadratic resonators and provides a route toward multicolour frequency combs, optical memories, and ultrafast photonic information processing.
\end{abstract}

\maketitle

\makeatletter
\begingroup
\def\@thefnmark{\ensuremath{}}
\@footnotetext{s.bogdanov@aston.ac.uk}
\endgroup
\makeatother

\section{Introduction}
Synchronization is one of the fundamental manifestations of nonlinear dynamics and underlies a broad range of physical systems, ranging from biological oscillators and chemical reactions to lasers \cite{strogatz2003sync,ROS01a}. Nonlinear optical systems provide an exceptionally rich environment for investigating synchronization phenomena, combining coherent wave interactions with precise control of propagation, dispersion, nonlinearity, and cavity dynamics. Among these platforms, $\chi^{(2)}$ nonlinear optical resonators are particularly attractive due to their efficient parametric frequency conversion, strong nonlinear coupling, and ability to support optical parametric oscillation, dissipative quadratic solitons, and broadband frequency combs through coherent coupling between the fundamental and second-harmonic fields (see e.g. \cite{PhysRevLett1961,dunn1999parametric,PRLMatsko,Leindecker:11,Marandi:12,Wang:18,He:19,Kivshar2020,Boes2023,Liu2023,Kivshar2003OpticalSF,KVandAM,Englebert2026NatureTopological} and references therein).
These advances have established quadratic microresonators as a versatile platform for nonlinear dynamics and ultrafast photonics \cite{He:19,Lu2025Photonic,sekine2025multi,Stokowski2024NatureOPO,Tang2024QuadraticComb}.

A fundamental challenge in quadratic nonlinear systems is the group-velocity mismatch between the interacting waves.
 In traveling-wave geometries, nonlinear parametric coupling can compensate for the linear walk-off, leading to velocity-locked quadratic solitons \cite{roy2022temporal}. In optical resonators, however, the repeated cavity boundary conditions introduce an additional synchronization constraint. 
 Consequently, previous studies have focused primarily on simultons and velocity-locked quadratic solitons, where the interacting fields propagate as a single localized structure with a common effective group velocity \cite{Buryak2002, Englebert2026QuadraticResonators}.

Here we identify a fundamentally different synchronization mechanism that does not rely on permanent temporal overlap or conventional velocity locking.
Instead of forming a single velocity-locked pulse pair, the resonator supports two synchronized pulse trains containing different integer numbers of pulses at the fundamental and second-harmonic frequencies.
When propagating over the cavity the pulse trains accumulate relative time delay, resulting in a mutual displacement by an integer number of pulse periods while reproducing the same global intracavity pattern. This phenomenon has been previously observed in an all-fiber laser cavity with a Raman amplifier \cite{nyushkov2023transient, turitsyn2025theory}.
Such behavior is analogous to the operation of an optical shift register and represents a previously unexplored synchronization topology in quadratic nonlinear resonators.

To reveal the physical origin of this phenomenon, we combine full cavity simulations with reduced analytical descriptions. We show that continuous parametric energy exchange produces a nonlinear correction to the relative pulse velocity, allowing the resulting walk-off to satisfy the rational shift-register condition, even when the corresponding linear criterion is not met. A minimal transport model obtained by neglecting dispersion and losses and a variational method-based analysis identify the origin of this correction and connect it quantitatively to the numerically observed synchronized states. The proposed mechanism, therefore, extends nonlinear synchronization beyond the conventional velocity locking associated with quadratic solitons: the interacting pulse trains need neither remain permanently overlapped nor contain identical numbers of pulses. Instead, continuous parametric energy exchange modifies the effective walk-off, allowing the system to self-organize into rational pulse-number states with a common repetition period. These results establish a new mechanism of nonlinear self-organization in quadratic resonators and provide the theoretical foundation for future experimental studies of multicolour pulse synchronization and frequency-comb generation.

\section{Theoretical models and materials}
\begin{figure}[t]
    \centering
    \includegraphics[width=\textwidth]{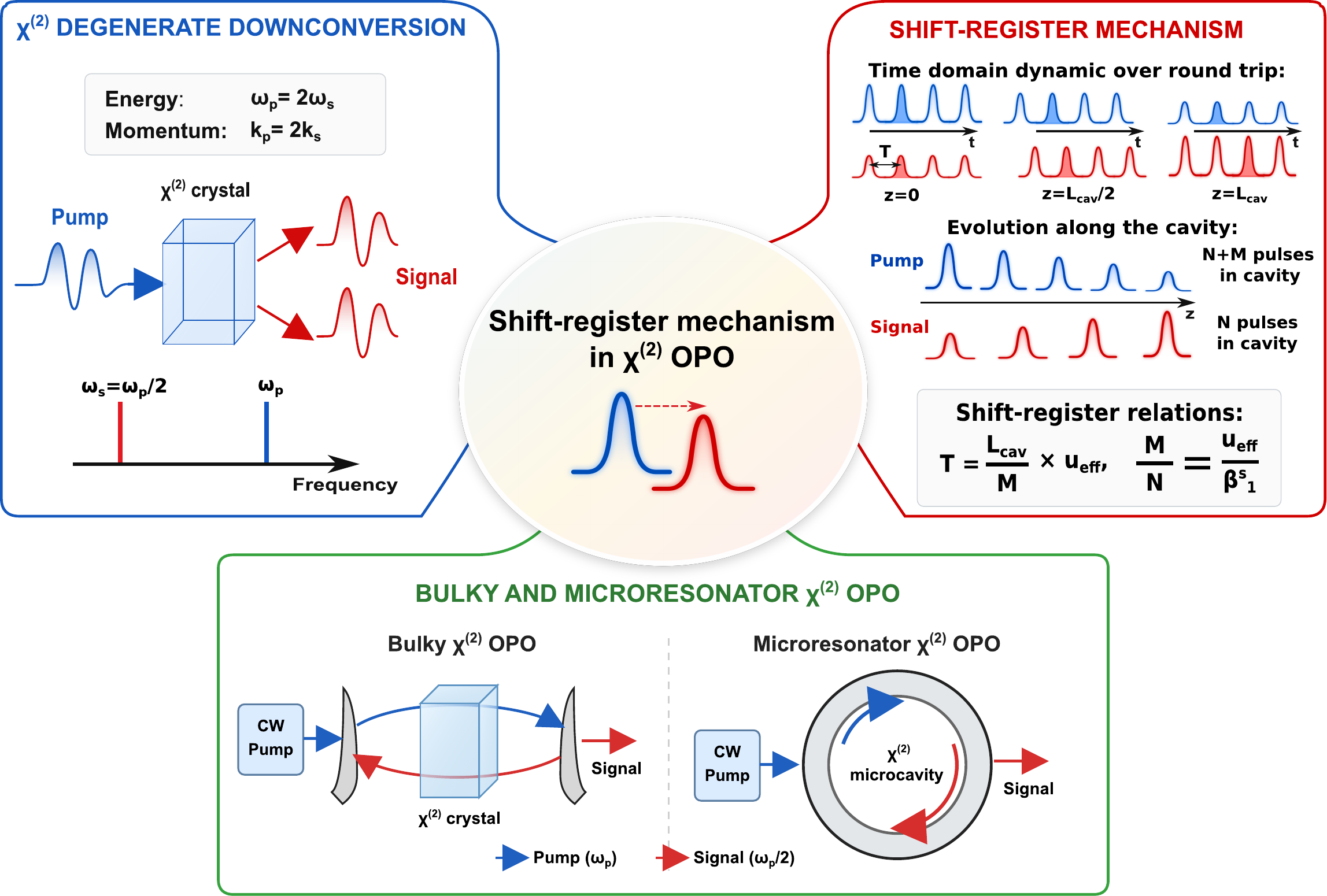}
    \caption{
Concept of shift-register synchronization in a degenerate
$\chi^{(2)}$ OPO. Quadratic downconversion couples the pump field $b$ at $2\omega$ to the half-harmonic signal field $a$ at $\omega$. Because the two fields propagate with different group velocities, their pulse trains acquire a relative temporal displacement during each cavity round trip. In a shift-register state, the complete intracavity pattern is reproduced after one round trip up to an integer temporal translation of one train relative to the other.
The mechanism can, in principle, be implemented in both bulk and integrated quadratic resonators.}
    \label{fig:concept_figure}
\end{figure}

In this section, we formulate the shift-register mechanism for a
doubly resonant degenerate $\chi^{(2)}$ optical parametric oscillator (OPO) 
(see Fig.~\ref{fig:concept_figure}).
This general formalism is applicable both to bulk and microresonators with a CW pump. We first introduce the coupled-field cavity model and then derive the synchronization conditions relating the cavity length, group-velocity mismatch, pulse repetition period, and integer numbers of circulating pulses. We finally discuss representative nonlinear materials and the experimentally accessible range of walk-off in degenerate OPOs. 

\subsection{Coupled-mode equations: Ikeda map formalism}

We describe nonlinear propagation in the quadratic waveguide using the coupled slowly varying envelope equations 
for the signal and pump envelopes, $a(z,t)$ and $b(z,t)$, respectively, following the notations of the experimental works \cite{hamerly2016reduced, Englebert2026QuadraticResonators}:
\begin{equation}
\begin{aligned}
    \frac{\partial a}{\partial z}
    &=
    \left(
        -\frac{\alpha^{(a)}}{2}
        - \mathrm{i}\,\frac{\beta^{(a)}_{2}}{2!}\,\frac{\partial^{2}}{\partial t^{2}}
        + \frac{\beta^{(a)}_{3}}{3!}\,\frac{\partial^{3}}{\partial t^{3}}
    \right)a
    + \kappa\,a^{\ast} b,
    \\
    \frac{\partial b}{\partial z}
    &=
    \left(
        -\frac{\alpha^{(b)}}{2}
        - u\,\frac{\partial}{\partial t}
        - \mathrm{i}\,\frac{\beta^{(b)}_{2}}{2!}\,\frac{\partial^{2}}{\partial t^{2}}
        + \frac{\beta^{(b)}_{3}}{3!}\,\frac{\partial^{3}}{\partial t^{3}}
    \right)b
    - \frac{1}{2} \kappa\,a^{2},
\end{aligned}
\label{eq:dual_envelope}
\end{equation}
where $\alpha^{(a,b)}$ denotes the linear loss coefficients for the signal and pump, respectively;
$\beta^{(a,b)}_{2}$ and $\beta^{(a,b)}_{3}$ are the group-velocity dispersion (GVD) and
third-order dispersion (TOD) coefficients, $u$ is the group-velocity mismatch (GVM) or walk-off
between pump and signal, and $\kappa$ is the effective quadratic nonlinear coupling
coefficient related to the second-order susceptibility and the overlap of the guided
modes \cite{Lu2025Photonic, hamerly2016reduced}.
The cavity dynamics are described using the Ikeda-map formalism:
propagation through the nonlinear crystal is followed by lumped
boundary conditions accounting for output coupling and other cavity
losses, phase detuning, and continuous-wave pump injection to the input of the nonlinear section \cite{nie2022direct}. Propagation
loss inside the nonlinear crystal is already included in Eq.~\eqref{eq:dual_envelope}.
The fields at the beginning of the $(n+1)$th round trip are therefore:

\begin{equation}
\begin{aligned}
    a^{(n+1)}(0,t)
    &= \sqrt{1-\theta^{(a)}} a^{(n)}(L_{\mathrm{cav}},t) e^{-i \delta^{(a)}},
    \\
    b^{(n+1)}(0,t)
    &= \sqrt{1-\theta^{(b)}} b^{(n)}(L_{\mathrm{cav}},t) e^{-i \delta^{(b)}} + \sqrt{\theta^{(b)}} b_{pump}.
\end{aligned}
\label{eq:boundary_conditions}
\end{equation}
where $\theta^{(a,b)}$ are the corresponding power-coupling
coefficients and $\delta^{(a,b)}$ are the cavity phase detunings.

\subsection{Shift-register mechanism}
We now formulate the shift-register mechanism (studied previously in Raman lasers \cite{nyushkov2023transient,turitsyn2025theory}) in the new context of a degenerate
$\chi^{(2)}$ OPO. Consider pump (main harmonic) and half-harmonic signal pulse trains
propagating in the resonator with group velocities $v_b$ and $v_a$,
respectively. 
The signal field $a$ has a higher group velocity than the pump field $b$, $v_a>v_b$, introducing a positive walk-off
$u=1/v_b-1/v_a>0$.
 
The temporal positions of pulses evolve according to $t= t_a(0) + \tau_a(z)=t_a(0)+ z/\nu_a$ and $t= t_b(0) +\tau_b(z) = t_b(0) +z/\nu_b$.
The round-trip times for the signal and pump harmonics are:
\begin{equation}\label{eq:roundtrip_time}
T_{a} = \frac{L_{cav}}{v_a},\;\;\;\; T_b = \frac{L_{cav}}{v_b} >T_a.
\end{equation}
Time delay (temporal shift) between two fields after one round trip is: $ \Delta T = T_b -  T_a$.
We are looking for solution with the same time period $T$ (interpulse interval) for signal and pump fields:
$a(\tau+T) = a(\tau), \;  b(\tau+T) = b(\tau)$
with  $N+M$ pulses in the microresonator
for the pump $b(\tau)$, and $N$ for the signal $a(\tau)$:
\begin{equation}\label{eq:num_pulses}
T_a = N \times T, \;\; T_b = (N + M) \times T.
\end{equation}

We seek synchronized solutions in which the two pulse trains in $a$ and $b$ reproduce the same intracavity pattern after every cavity round trip, while undergoing a deterministic relative temporal shift of $M$  pulse periods, in other words, $\Delta T = M T$.
Substituting Eq.~\eqref{eq:num_pulses} into Eq.~\eqref{eq:roundtrip_time} gives two relations. The first synchronization condition determines the pulse repetition period through the effective walk-off,
\begin{equation}\label{eq:T}
     T =\frac{L_{\mathrm{cav}}}{M} \times u_{eff},\;  \;\;\; u_{eff} = u_{lin} + \Delta u_{NL}.
\end{equation}
Here $u_{lin} = \beta_1^{(b)} - \beta_1^{(a)}$ is the linear group velocity mismatch, and  $\Delta u_{NL}$ is produced by parametric gain/depletion and asymmetric energy exchange between pulses. 
Importantly, shift-register synchronization does not require
$u_{\rm eff}=0$. Instead, the effective walk-off must accumulate to
an integer number of pulse periods over one round trip, $L_{\rm cav}u_{\rm eff}=MT$.
This distinguishes the present mechanism from conventional
velocity locking.
The second synchronization condition relates the numbers of pulses circulating at the two harmonically related frequencies. 
 
\begin{equation}\label{eq:N}
   1+ \frac{M}{N} = \frac{v_a}{v_b}.
\end{equation}

In the linear case $v_{a}=1/ \beta_1^{(a)}$, $  v_{b}=1/\beta_1^{(b)}$ and conditions read:
\begin{equation}\label{eq:T1}
    T = \frac{L_{\mathrm{cav}}}{M} \times u_{lin},
\end{equation}
\begin{equation}\label{eq:N1}
   1+ \frac{M}{N} = \frac{\beta_1^{(b)} }{\beta_1^{(a)} }.
\end{equation}

There are several design considerations based on Eq.~\eqref{eq:T} and Eq.~\eqref{eq:N}. 
In the purely linear case, Eq.~\eqref{eq:N1} requires the ratio of the two round-trip velocities to coincide exactly with a rational pulse-number
ratio, which is generally restrictive. Nonlinear parametric interaction
relaxes this requirement by modifying the effective relative velocity.
Unlike conventional soliton trapping, however, the nonlinear correction
does not force the two fields to propagate at the same velocity.
Instead, it renormalizes the accumulated walk-off so that the discrete
condition $L_{\rm cav}u_{\rm eff}=MT$ is satisfied.

In the case of asynchronous pumping, with the period on pump $T$ being a tunable parameter and fixed group-velocity mismatch $u$, cavity length $L_{cav}$ should be an integer number ($M$) of $T/u_{eff}$. For the fixed $L_{cav}$, the period should be chosen as in Eq.~\eqref{eq:T}. In the case of pulse formation from the CW pump, the resonator length determines the joint period $T$ of the signal and pump pulse trains in the cavity. 

Eq.~\eqref{eq:T} and Eq.~\eqref{eq:N} constitute the fundamental design equations of the shift-register mechanism. Eq.~\eqref{eq:T} determines the pulse repetition period through the effective walk-off, while Eq.~\eqref{eq:N} specifies the rational pulse-number relation required for synchronized operation.

\subsection{Materials}
\begin{figure}[h!]
    \centering
    \includegraphics[width=0.6\textwidth]{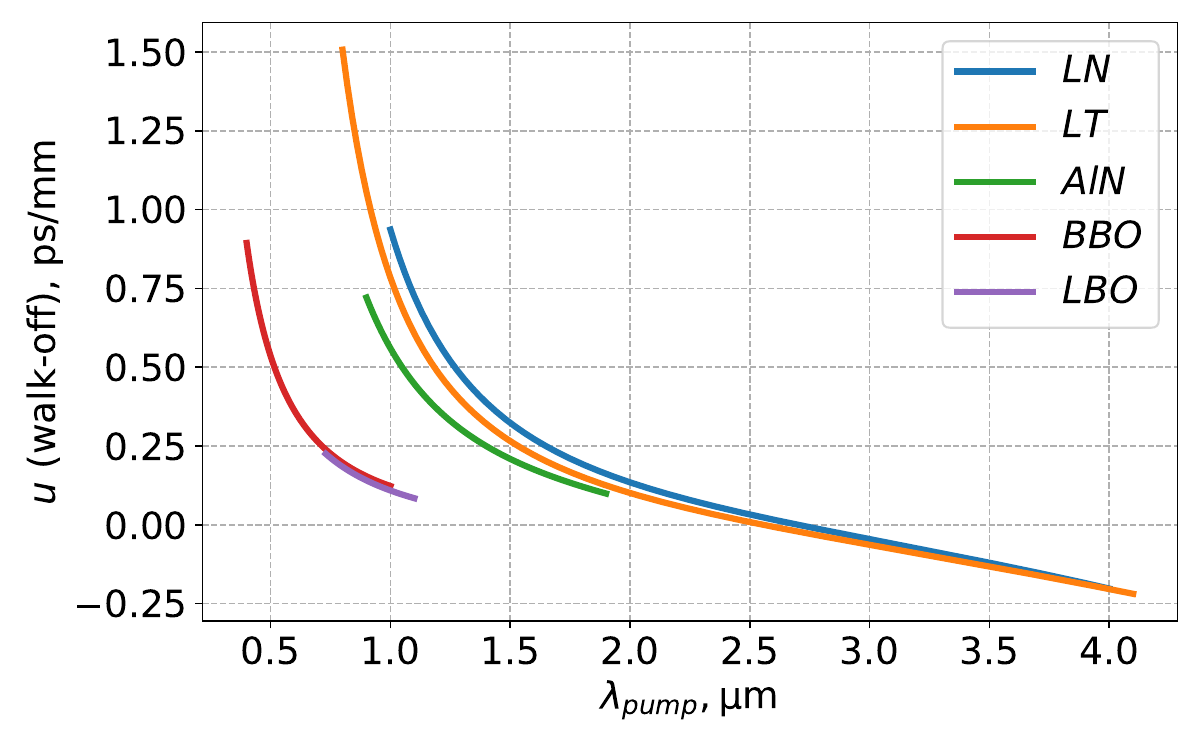}
    \caption{
Linear group-velocity mismatch (walk-off) for representative
$\chi^{(2)}$ materials (including lithium niobate (LN), lithium tantalate (LT), aluminum nitride (AlN), barium borate (BBO), and lithium triborate (LBO)) in degenerate downconversion as a function of pump wavelength, calculated from published Sellmeier relations. The accessible values span the range required for the shift-register states investigated below, showing that the synchronization conditions are compatible with realistic bulk and integrated quadratic platforms.}
    \label{fig:fig_woff}
\end{figure}

Typical nonlinear materials used to generate $\chi^2$ nonlinearity, and suitable for implementing the proposed mechanism,  such as lithium niobate (LN), lithium tantalate (LT), aluminum nitride (AlN), barium borate (BBO), and lithium triborate (LBO) among many others, are widely adopted in bulk OPO and microresonators
\cite{dunn1999parametric, Lu2025Photonic}.
For these materials, the walk-off as a function of pump wavelength in the degenerate downconversion process is shown in Fig.~\ref{fig:fig_woff}. These values were calculated with the Sellmeier equations and coefficients adopted from the literature \cite{gayer2008temperature, bruner2003temperature, bowman2018optical,
velsko1991phase,
eimerl1987optical}.

\section{Shift-register mechanism in a doubly resonant OPO with CW pump}

\begin{table}[!t]
    \centering
    \caption{Parameters used in the coupled-envelope simulation of the PPLN waveguide OPO, based on \cite{hamerly2016reduced}.}
    \label{tab:params}
    \begin{tabular}{llll}
        \hline
        Symbol & Description & Value & Unit \\
        \hline
        $\lambda^{(a)}, \lambda^{(b)}$ & Signal, pump wavelength & $1500$, $750$ & nm \\        $\alpha^{(a)}_{2},\alpha^{(b)}_{2}$ & Waveguide Loss & $0.00691$ & mm$^{-1}$ \\
        $\beta^{(a)}_{2}, \beta^{(b)}_{2}$ & Signal and pump GVD & $1.12\times 10^{-4}$, $4.06\times 10^{-4}$ & ps$^{2}$/mm \\
        $\beta^{(a)}_{3}, \beta^{(b)}_{3}$ & Signal and pump TOD & $3.09\times 10^{-5}$, $2.51\times 10^{-5}$ & ps$^{3}$/mm \\\
        $\kappa$ & Nonlinearity & $5.16\times 10^{-5}$ & ps$^{1/2}$/mm \\
        $b_0$ & Pump normalization amplitude & $3.84 \times 10^2$ & ps$^{-1/2}$ \\
        $k_a^{out}$ & Signal output coupling &
        0.7 & \\
        $k_b^{out}$ & Pump output coupling &
        0.01 & \\
        \hline
    \end{tabular}
\end{table}

\begin{figure}[!t]
    \centering
    \includegraphics[width=\textwidth]{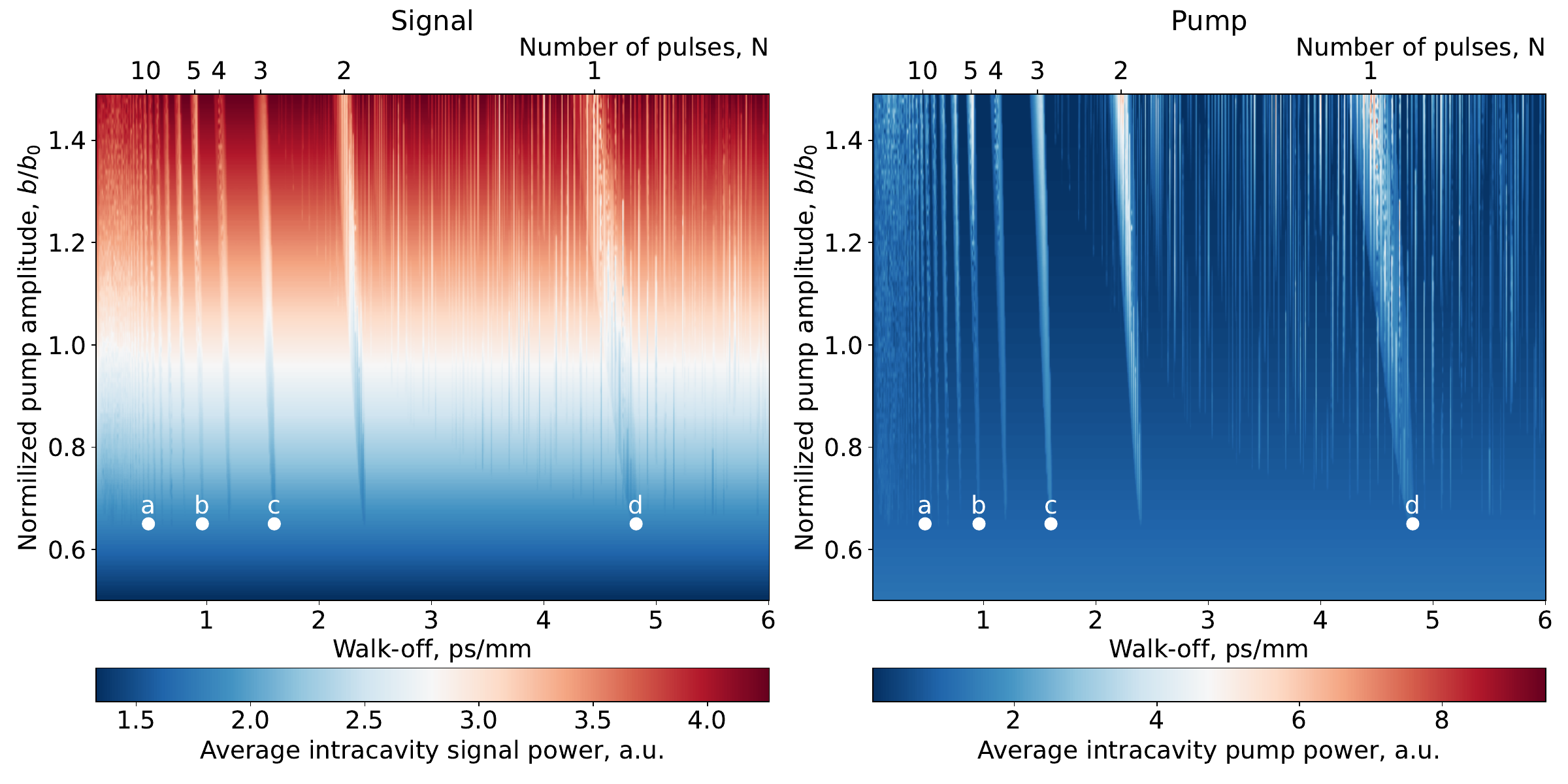}
    \caption{
Numerical synchronization maps obtained by adiabatically varying the normalized input pump amplitude $b/b_0$ for each value of the linear walk-off $u$. Shown are the intracavity signal (left) and pump (right) powers. Narrow high-power domains correspond to stable shift-register states with different integer numbers $N$ of signal pulses in the cavity and $N+M$ pump pulses. The systematic shift of these domains with pump amplitude demonstrates that the synchronization condition cannot be determined by the linear walk-off alone and indicates a nonlinear renormalization of the effective relative velocity. White markers ``a'', ``b'', ``c'', ``d'' with corresponding $N$ values 10, 5, 3, 1 indicate representative states shown in Fig.~\ref{fig:fig_4}.} 
        \label{fig:fig_3}
\end{figure}

We perform numerical simulations of a doubly resonant degenerate downconversion OPO driven by a CW pump. The system described by Eq.~\eqref{eq:dual_envelope} is solved using the split-step Fourier method \cite{agrawal2000nonlinear} with the boundary conditions given by Eq.~\eqref{eq:boundary_conditions}. Unless otherwise stated, the parameters are taken from Ref.~\cite{hamerly2016reduced}. In our analysis, however, the walk-off $u$ is treated as a free parameter within a given range, which can be achieved by tuning the pump wavelength and selecting an appropriate crystal material, as illustrated in Fig.~\ref{fig:fig_woff}. The crystal length is also varied between 10 and 100~mm to cover the different operating regimes considered. For the doubly resonant configuration, the signal and pump output coupling coefficients are set to $k_a^{out}=0.7$ and $k_b^{out}=0.01$ correspondingly. The complete set of parameters is summarized in Table~\ref{tab:params}.

\begin{figure}[t]
    \centering
    \includegraphics[width=\textwidth]{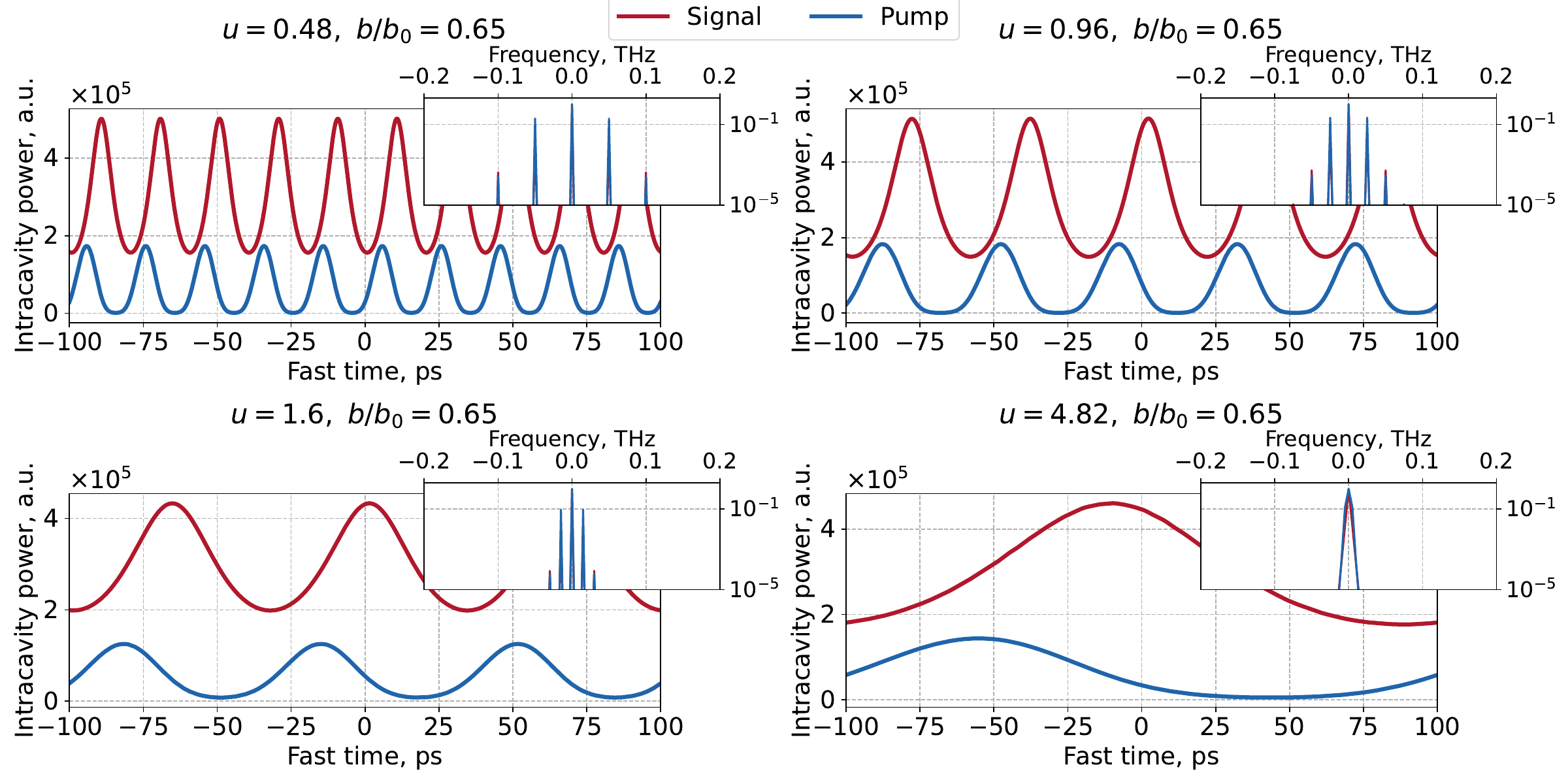}
    \caption{
    Representative stationary shift-register states selected from the synchronization domains in Fig.~\ref{fig:fig_3}, corresponding from upper left to lower right to $N=10,5,3,$ and $1$ pulses per cavity round trip. Main panels show the intracavity signal and pump powers versus fast time; insets show the corresponding normalized spectra. As the walk-off increases, the synchronized state reorganizes into progressively smaller integer pulse-number states while preserving regular pulse-train operation.}
      \label{fig:fig_4}
\end{figure}

In the first numerical experiment, we fix the crystal length at 40~mm and sweep the walk-off $u$ from 0.1 to 6~ps/mm with a step of 0.005~ps/mm. For each value of $u$, the simulation is initialized below the generation threshold at $b/b_0=0.5$ and propagated for $10,000$ round trips. The normalized pump amplitude is then increased in increments of 0.01, with the field propagated for another $10,000$ round trips at each step, covering the range $b/b_0=0.5$--1.49 in 100 steps. At each pump level, the final signal-field configuration from the preceding step is used as the initial condition, except for the first step, where a weak CW signal is introduced. This continuation procedure is commonly referred to as the adiabatic procedure \cite{smirnov2021soliton}.

\begin{figure}[t]
    \centering
    \includegraphics[width=\textwidth]{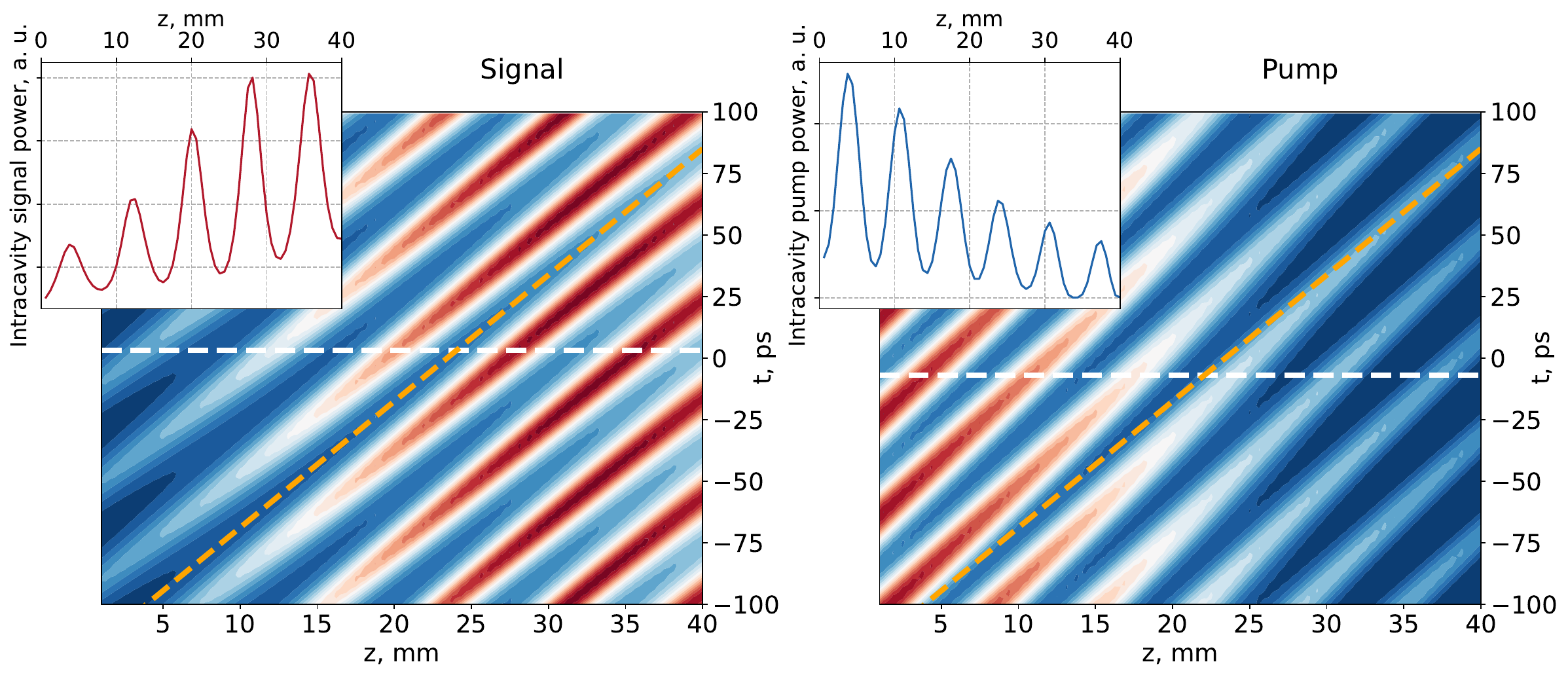}
\caption{Intracavity spatiotemporal evolution of the signal (left) and pump (right) in a representative shift-register state at $u=0.9~\mathrm{ps/mm}$ and $b/b_0=1.05$. The diagonal pulse trajectories reveal the different effective propagation velocities of the two harmonics (as indicated by their different slopes; the orange line serves to demonstrate that having the same tilt in both plots). Nevertheless, the two pulse trains remain globally
synchronized: the cavity contains $N=5$ signal pulses and $N+M=6$ pump pulses, so that their relative displacement accumulated during one round trip corresponds to one pulse period ($M=1$). Insets show
cross-sections along the indicated dashed lines.}
     \label{fig:fig_5}
\end{figure}

Fig.~\ref{fig:fig_3} reveals a sequence of distinct synchronization domains, visible as bright bands in the $(u,b/b_0)$ plane, each corresponding to a stable integer pulse-number state. These bands are tilted, indicating that increasing the pump strength shifts the synchronization regions toward smaller values of the imposed linear walk-off. This trend provides the direct indication that the parametric interaction introduces a pump-dependent nonlinear correction, $\Delta u_{\rm NL}$, to the effective walk-off in Eq.~\eqref{eq:T}.
The figure also demonstrates that for the small walk-off values, the dynamics are irregular, and no well-defined pulse-generation domains are observed. As the walk-off increases, distinct synchronization regions emerge and become increasingly well separated.

Fig.~\ref{fig:fig_4} depicts states corresponding to different number of signal pulses in the cavity, $N=\mathrm{10}$, 5, 3 and 1 (also marked with ``a'', ``b'', ``c'', ``d'' in Fig.~\ref{fig:fig_3}). 
When the shift-register condition is satisfied at the specific walk-off value the signal and pump envelopes shows stable pulse generation. Therefore, the number of pulses $N$ is dictated by the chosen resonant value of the walk-off.
This explains the bright bands in Fig.~\ref{fig:fig_3} and the different numbers of pulses observed in the corresponding regimes.

Fig.~\ref{fig:fig_5} shows the intracavity spatiotemporal evolution of a representative $M=1$ shift-register state. A cross-section at a fixed time reveals $N+M=6$ pump pulses and $N=5$ signal pulses circulating in the microresonator. The unequal pulse numbers arise from the group-velocity mismatch between the interacting waves and are a characteristic feature of the shift-register regime. Accordingly, the signal and pump trajectories have different slopes, indicating that the two fields are not locally velocity locked. Nevertheless, over one round trip, their relative temporal displacement is exactly one pulse period, while the cavity accommodates $N=5$ signal pulses and $N+M=6$ pump pulses. Thus, the figure directly illustrates the distinction between shift-register synchronization, which is established only after a complete round trip, and conventional velocity locking, which requires local matching of the pulse velocities.
 
\begin{figure}[!t]
    \centering
    \includegraphics[width=0.6\linewidth]{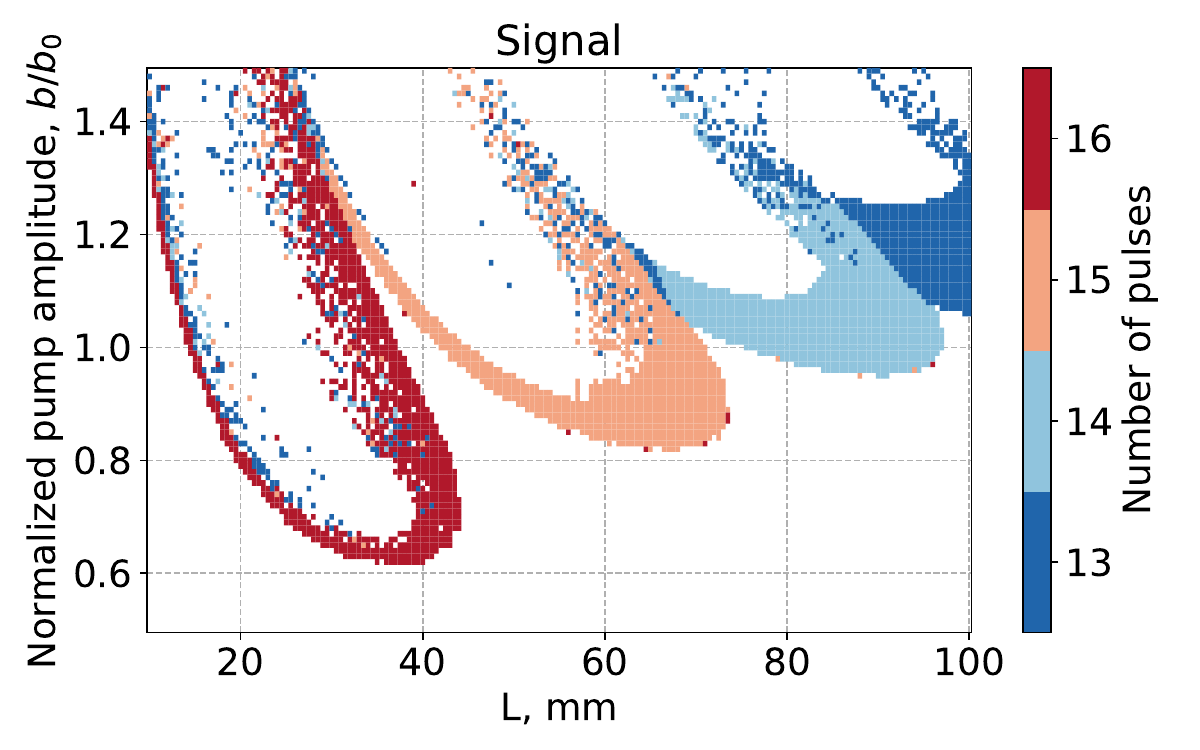}
    \caption{
Stable shift-register domains as a function of nonlinear interaction length $L$ and normalized input pump amplitude $b/b_0$ for fixed walk-off $u=0.3~\mathrm{ps/mm}$. Colours denote the number $N$ of signal pulses identified in the converged intracavity state; white regions correspond to states rejected by the stability and pulse-contrast criteria. As $L$ increases, the accumulated walk-off grows and drives successive transitions between neighboring integer pulse-number states, consistent with the synchronization condition of Eq.~\eqref{eq:T}.}
    \label{fig:fig_6}
\end{figure}

In the second numerical experiment, we fix the walk-off at $u=0.3$ and vary the crystal length $L$ from 10 to 100~mm in 1~mm increments, using the same adiabatic procedure as above. Depending on $L$, we observe different numbers of pulses circulating in the cavity, ranging from 13 to 16, corresponding to the distinct colored domains in Fig.~\ref{fig:fig_6}. To plot the map, we automatically determine the number of pulses in each signal envelope and identify stable pulse-train states based on two criteria. First, the coefficient of variation of the pulse peak powers, defined as the ratio of their standard deviation to their mean, must satisfy $\mathrm{CV}_{\mathrm{peak}}<0.15$, excluding states with significant pulse-to-pulse fluctuations. Second, we require a minimum contrast between the pulse peaks and the background to distinguish genuine pulse trains from CW states with noise-induced peaks.

For a given $L$, stable pulse generation in the shift-register regime emerges as the pump power increases beyond the generation threshold. Further growth of the pump power eventually drives the cavity into a chaotic state, destroying the regular pulse train with a fixed number of pulses $N$ and giving rise to white regions within the colored domains.
One more remarkable feature is that the pulse spacing $T$ increases with the crystal length, in agreement with the prediction of Eq.~\eqref{eq:T}. Beyond a critical length, however, the system transitions to a state with one fewer pulse, accompanied by an increase in the generation threshold. This transition cannot be explained by the linear synchronization conditions given by Eqs.~\eqref{eq:T1} and \eqref{eq:N1}, indicating that nonlinear effects modify the effective walk-off and, consequently, the pulse spacing and number of pulses, as described by Eqs.~\eqref{eq:T} and \eqref{eq:N}.

\subsection{Origin of the nonlinear walk-off}
To identify the physical origin of the nonlinear correction
$\Delta u_{\rm NL}$ inferred from Fig.~\ref{fig:fig_3} -- Fig.~\ref{fig:fig_6}, we first consider a
minimal transport model obtained by neglecting chromatic dispersion in Eq.~\eqref{eq:dual_envelope}. This approximation reduces the governing equations to a pair of coupled transport equations describing the evolution of the pump and signal envelopes:

\begin{equation}
\begin{aligned}
    \frac{\partial a}{\partial z} + \beta^{(a)}_{1}\,\frac{\partial a}{\partial t}
    &=
     -\frac{\alpha^{(a)}}{2}a
    + \kappa\,a^{\ast} b,
    \\
    \frac{\partial b}{\partial z} + \beta^{(b)}_{1}\,\frac{\partial b}{\partial t}
    &=
        -\frac{\alpha^{(b)}}{2}b
    - \frac{1}{2} \kappa\,a^{2}.
\end{aligned}
\label{eq:transport_envelope}
\end{equation}

Eq.~\eqref{eq:transport_envelope} can be used to examine the nonlinear contribution to the walk-off. 
Assuming well-separated pump and signal pulse trains whose pulses translate coherently (moving as a whole) within each train, we characterize their effective motion through the temporal centroids
(first moments) - integral characteristics of the corresponding speeds:
\begin{equation} \label{eq:RMS}
    T_{c}^{(a)}= \frac{\int{t |a(z,t)|^2 dt}}{\int{|a(z,t)|^2 dt}},\;\;\;T_{c}^{(b)} = \frac{\int{t |b(z,t)|^2 dt}}{\int{|b(z,t)|^2 dt}}.
\end{equation}
After straightforward manipulation, defining
$R(t)=\mathrm{Re}[a^2b^*]$, we obtain the effective centroid walk-off (more details are in Supplement 1):
\begin{equation} 
\label{eq:U_eff}
    u_{c} =\frac{d T_{c}^{(b)}}{dz}- \frac{d T_{c}^{(a)}}{dz}= \beta^{(b)}_{1} - \beta^{(a)}_{1}  - 2 \kappa \; \frac{\int{(t-T_{c}^{(a)}) R(t)dt}}{\int{|a(z,t)|^2 dt}}  - \kappa \; \frac{\int{(t -T_{c}^{(b)} ) R(t)dt}}{\int{|b(z,t)|^2 dt}}. 
\end{equation}

\begin{figure}[t]
    \centering
    \includegraphics[width=\textwidth]{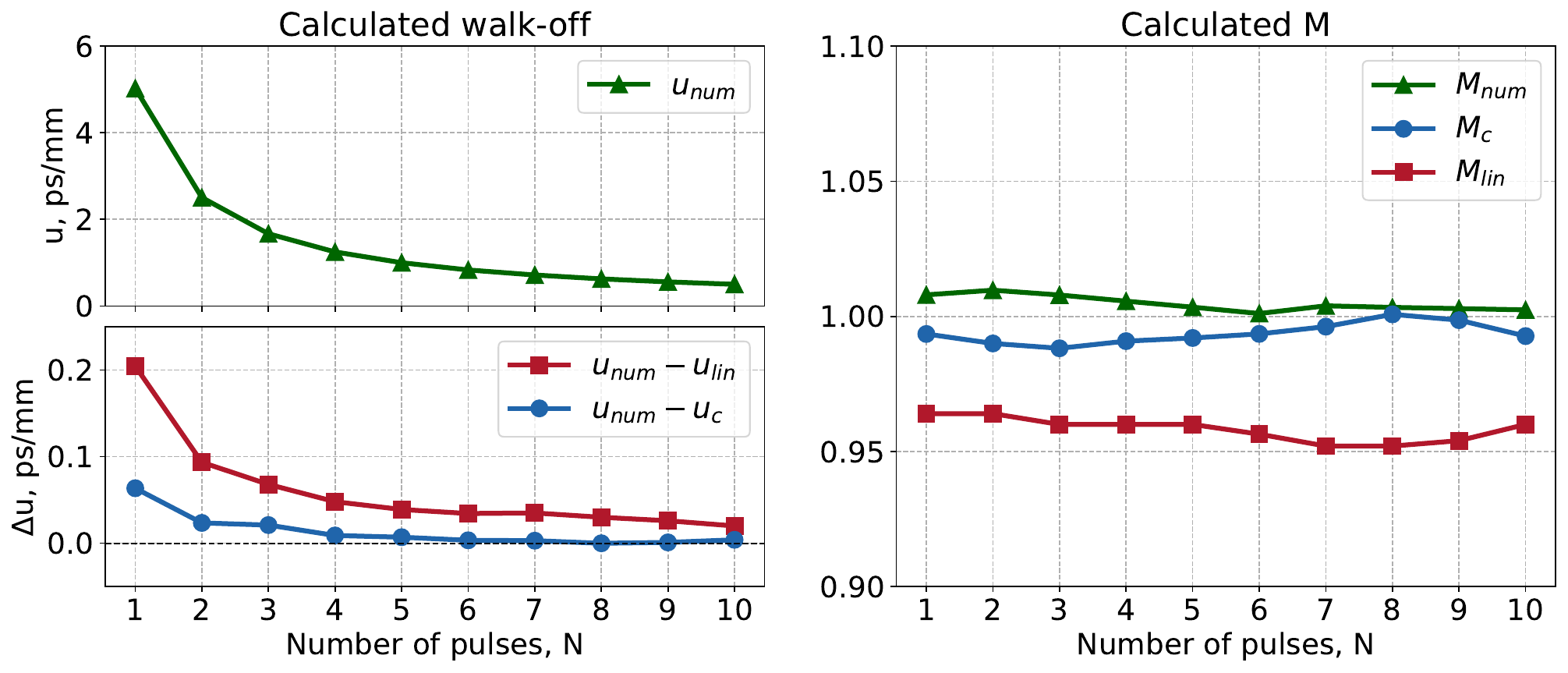}
    \caption{Calculated walk-off mismatches and shift index $M$ for states with different signal pulse numbers $N$. The linear prediction based on $u_{\rm lin}$ lies farther from the synchronization condition, whereas the nonlinear centroid correction from Eq.~\eqref{eq:U_eff} brings the effective walk-off into close agreement with it, yielding $M\approx1$, consistent with the full dispersive simulations.}
      \label{fig:fig_u_eff}
\end{figure}

Fig.~\ref{fig:fig_u_eff} illustrates how the nonlinear interaction renormalizes the linear walk-off, $u_{lin}$, toward the shift-register condition. For states with different numbers $N$ of signal pulses, we calculate the walk-off directly from numerical simulations of the full model, $u_{num}$, and estimate $u_{c}$ with Eq.~\eqref{eq:U_eff}.
The latter is closer to the full-model values than the linear walk-off $u_{lin}$: $u_{num}-u_{c}$ is smaller than $u_{num}-u_{lin}$ for all $N$. This confirms that nonlinearity considered in the transport model is the key contributor to the walk-off corrections, $\Delta u_{\mathrm{NL}}$. The same trend is observed for the shift indices $M_{lin}$, $M_{c}$, $M_{num}$ calculated from $u_{lin}$, $u_c$, and $u_{num}$ with Eqs.~\eqref{eq:N} and \eqref{eq:N1}. As expected, $M_{num}$ remains closer to unity for all $N$ since $M$ takes integer values and only regimes with $M=1$ are observed in our analysis. Although $M_{lin}$ is farther from the true value, the estimates $M_c$ show better agreement, thereby justifying the transport model. The variation of $M_{lin}$, $M_c$, and $M_{num}$ in Fig.~\ref{fig:fig_u_eff} (right) and the deviation $M_{num}$ from exact unity is explained by the complex structures of the bright bands (see Fig.~\ref{fig:fig_3}). In our analysis, we scan the $(u,b/b_0)$ plane using a finite-step grid and select shift-register regimes closest to the generation threshold, which can yield operating states with distinct properties for different $N$.

\subsection{Reduced variational description}

\begin{figure}[!t]
    \centering
    \includegraphics[width=\textwidth]{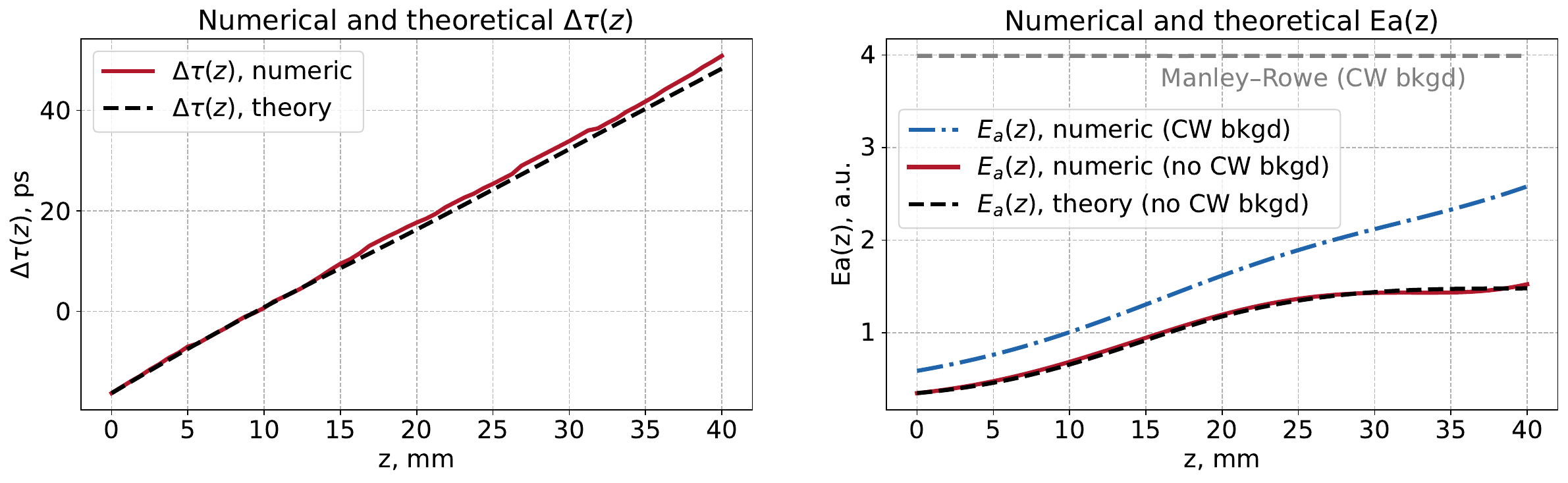}
    \caption{
    Comparison of the $\Delta \tau (z)$ and $E_a(z)$ numerically calculated with the full model and theoretical values from Eq.~\eqref{eq:variational_delay} and Eq.~\eqref{eq:variational_energy}.
    For $E_a(z)$, the blue dash-dotted curve shows the total signal energy obtained from the full model with compensated intracavity loss (multiplied by $e^{\alpha z}$). The red solid curve depicts the numerical pulse energy after subtracting the CW background and applying similar intracavity loss compensation. The latter closely follows the variational result, consistent with the localized-pulse ansatz used in deriving Eq.~\eqref{eq:variational_energy}.}
    \label{fig:dtau_Ea_theory}
\end{figure}

A complementary collective-coordinate analysis provides a microscopic
interpretation of the nonlinear walk-off correction. The full derivation,
given in Supplement 1, represents the two fields by localized periodic
pulse trains and reduces the coupled-wave equations to evolution equations
for the pulse energies, relative temporal position, frequency mismatch,
and relative phase. A particularly simple description is obtained by approximating both the signal and pump pulses by Gaussian profiles. 
In general, the relative temporal coordinate obeys
\begin{equation}
\frac{d\Delta\tau}{dz}
=
u+\Delta u_{\mathrm{disp}}+\Delta u_{\mathrm{NL}},
\label{eq:variational_walkoff_general}
\end{equation}
where $\Delta u_{\mathrm{NL}}$ is determined by the nonlinear overlap
between the reference signal pulse and the pump pulses encountered during
propagation and $\Delta u_{\mathrm{disp}}$ is the contribution due to dispersion.

To obtain a simple analytical description of the numerically observed
states, we consider the regime of small frequency mismatch
$Q=T_s(2\Omega_a-\Omega_b)$ and relative nonlinear phase
$\Phi=2\phi_a-\phi_b$. In this limit, the reduced dynamics simplify
to two coupled equations for the
relative temporal displacement $\Delta\tau$ and the signal-pulse energy $E_a$:
\begin{equation}
\frac{d\Delta\tau}{dz}
=
u+
\frac{\Gamma\left(3E_a-N_{\rm MR}\right)}
{3\sqrt{\left(N_{\rm MR}-E_a\right)/2}}
\sum_{n=-\infty}^{\infty}
\left(\Delta\tau+nT\right)F_n ,
\label{eq:variational_delay}
\end{equation}
and
\begin{equation}
\frac{dE_a}{dz}
=
\Gamma E_a
\sqrt{2\left(N_{\rm MR}-E_a\right)}
\sum_{n=-\infty}^{\infty}F_n ,
\label{eq:variational_energy}
\end{equation}
where $\Gamma$ is the effective quadratic coupling coefficient
defined in Supplement 1,
\begin{equation}
F_n=
\sqrt{\frac{2\pi}{3}}
\exp\left[
-\frac{\left(\Delta\tau+nT\right)^2}{3T_s^2}
\right],
\label{eq:gaussian_overlap}
\end{equation}
and $N_{\rm MR}=E_a+2E_b$
is the Manley--Rowe invariant.

Eqs.~\eqref{eq:variational_delay} and \eqref{eq:variational_energy} reveal the physical mechanism particularly
clearly. As a signal pulse slips through the pump train, it undergoes a
sequence of exponentially localized nonlinear encounters indexed by $n$.
The same encounters that redistribute energy between the two harmonics
also modify their relative temporal motion. The nonlinear interaction
therefore does not need to cancel the walk-off locally; instead, it
renormalizes the accumulated relative displacement so that the
shift-register condition
\begin{equation}
\Delta\tau(L)=\Delta\tau(0)+MT
\end{equation}
can be satisfied.

Fig.~\ref{fig:dtau_Ea_theory} depicts $\Delta \tau(z)$ and $E_a(z)$ obtained from the full numerical model, Eq.~\eqref{eq:dual_envelope}, and the reduced variational description. 
The $\Delta \tau(z)$ calculated with Eq.~\eqref{eq:variational_delay} is consistent with the direct observation of the pulses displacement in the cavity. For the signal pulse energy, $E_a(z)$, the reduced variational ansatz describes the localized pulsed component of the intracavity fields and does not include the weak continuous-wave background present in the full cavity simulations. Accordingly, the total numerical signal energy contains an additional background contribution that becomes increasingly visible toward the end of the nonlinear section. When this CW contribution is subtracted from the numerical field, the resulting pulse energy shows close agreement with the reduced variational prediction over the entire propagation interval.

The complete variational derivation, including the more general four-dimensional reduced system for
$(E_a,Q,\Delta\tau,\Phi)$, is provided in Supplement 1.

\section{Discussion and outlook}
In this work, we introduced a shift-register synchronization regime
in optical resonators with quadratic nonlinearity, in which harmonically
related pulse trains containing different numbers of pulses reproduce
the same intracavity state after each round trip, up to a deterministic
discrete temporal translation.
This is fundamentally different from the velocity-locking mechanisms previously studied in quadratic nonlinear systems. Conventional quadratic conservative and dissipative solitons rely on nonlinear compensation of group-velocity mismatch to form a single composite structure propagating with a common velocity. In contrast, the synchronized states reported here consist of two pulse trains containing different integer numbers of pulses that reproduce the same intracavity pattern after every resonator round trip while undergoing a deterministic relative temporal translation. Synchronization therefore occurs at the level of the global cavity dynamics rather than through permanent temporal overlap of individual pulses.

This distinction provides a different perspective on synchronization in nonlinear optical resonators. Instead of requiring complete velocity matching, the resonator supports synchronized solutions satisfying rational relationships between the pulse repetition period, the cavity round-trip time, and the effective group-velocity mismatch. Continuous parametric interaction modifies the effective walk-off sufficiently for these rational synchronization conditions to be fulfilled over a finite parameter range. The nonlinear interaction therefore acts as an adaptive mechanism that enables the existence of shift-register states, rather than constituting the phenomenon itself.

The theoretical analysis further shows that the essential physics is captured by a remarkably simple transport model. Although this reduced description neglects chromatic dispersion, it reproduces the pulse repetition period and predicts the nonlinear correction to the effective walk-off with good agreement with the complete cavity simulations. This observation indicates that the formation of shift-register states is governed primarily by the interplay between nonlinear energy exchange and group-velocity mismatch, whereas higher-order dispersive effects mainly influence the detailed pulse shape and stability. Such separation of physical mechanisms is encouraging because it suggests that the synchronization principle is considerably more general than the specific model investigated here.

An attractive feature of the proposed mechanism is that it is formulated through simple synchronization conditions involving only cavity length, group velocities, and integer pulse numbers. These relations provide practical design rules for engineering synchronized multicolour pulse trains in quadratic resonators. Since the required walk-off can be controlled through material selection, wavelength tuning, waveguide geometry, or cavity parameters, the shift-register regime should be accessible in a broad range of integrated and bulk $\chi^{(2)}$ platforms.

The concept may also extend beyond degenerate optical parametric oscillators. Similar synchronization principles may emerge in other nonlinear resonators supporting coupled waves with different group velocities, including non-degenerate parametric oscillators, sum- and difference-frequency generation, coupled microcomb systems, and hybrid nonlinear photonic platforms. More generally, the underlying idea of synchronization through rational temporal translations may represent a generic mechanism in periodically driven nonlinear systems where interacting waves repeatedly experience both distributed nonlinear evolution and discrete cavity boundary conditions.

Several directions remain open for future investigation. A natural extension is the analytical characterization of the complete family of shift-register states, including their existence, stability, and bifurcation structure. Reduced-order descriptions based on variational methods or other collective-coordinate approaches may provide additional insight into the nonlinear dynamics near synchronization thresholds. It will also be important to investigate the influence of higher-order dispersion, noise, detuning, and imperfect phase matching on the robustness of the synchronized states.

An important next step is experimental verification.
Recent advances in integrated lithium niobate and other quadratic microresonator platforms provide the necessary combination of strong nonlinear interaction, low loss, and dispersion engineering to explore the predicted synchronization regime. Experimental observation of deterministic shift-register dynamics would not only validate the present theoretical predictions but would also establish a new mechanism for generating synchronized multicolour pulse trains with controllable temporal structure.

In summary, the shift-register regime expands the current understanding of nonlinear synchronization in optical resonators. Rather than synchronizing individual pulses through complete velocity locking, the resonator synchronizes the evolution of two entire pulse trains through discrete temporal translations that repeat from one cavity round trip to the next. This introduces a new class of synchronized nonlinear states and opens opportunities for both fundamental studies of nonlinear dynamics and practical implementations in ultrafast photonics.

\section{Conclusion}
We have demonstrated a shift-register synchronization regime in quadratic nonlinear resonators in which harmonically related pulse trains containing different integer numbers of pulses reproduce the same global intracavity state after each round trip up to a discrete temporal translation. Full cavity simulations, centroid analysis, and a reduced variational description show that continuous parametric
energy exchange renormalizes the effective walk-off, enabling the rational synchronization condition to be satisfied beyond the linear matching regime. These results establish synchronization through periodic self-reproduction of entire pulse trains as a distinct
mechanism from conventional velocity locking and provide a route toward experimentally realizable multicolour synchronized states in
quadratic resonators.
We anticipate that these results will stimulate further theoretical
investigations of nonlinear synchronization in resonators and provide a foundation for experimental studies of multicolour pulse dynamics
in integrated quadratic photonic platforms.

\section{Acknowledgment}
This work was supported by Engineering and Physical Sciences Research Council (EP/W002868/1). A.E.B. acknowledges support of the Russian Science Foundation (№ 24-12-00314, https://rscf.ru/en/ project/24-12-00314/).

\bibliography{References_SS}

\end{document}


\renewcommand{\theequation}{S\arabic{equation}}

\title{Shift-register synchronization of dual-wavelength pulse trains in resonators with quadratic nonlinearity: supplemental}

\author{Stepan~Bogdanov}
\thanks{These authors contributed equally to this work.}
\affiliation{\mbox{Aston Institute of Photonic Technologies, Aston University, Birmingham, B4 7ET, UK}}

\author{Anastasiya~Bednyakova}
\thanks{These authors contributed equally to this work.}
\affiliation{Novosibirsk State University,
Novosibirsk, 630090, Russia}

\author{Sergei~K.~Turitsyn}
\affiliation{\mbox{Aston Institute of Photonic Technologies, Aston University, Birmingham, B4 7ET, UK}}

\begin{abstract} 
The Supplemental Document provides further details for the manuscript ``Shift-register synchronization of
dual-wavelength pulse trains in resonators with quadratic nonlinearity''. First, we examine the nonlinear correction to the effective walk-off using a reduced transport model. We then develop a variational description of periodic pulse trains that complements the full numerical model and provides physical insight into the role of quadratic interaction in renormalizing the walk-off and establishing shift-register synchronization.
\end{abstract}

\maketitle

\makeatletter
\begingroup
\def\@thefnmark{\ensuremath{}}
\@footnotetext{s.bogdanov@aston.ac.uk}
\endgroup
\makeatother

\section{Walk-off analysis using transport equations}
To identify the primary physical mechanisms responsible for the nonlinear modification of walk-off and the emergence of the
shift-register regime, we consider a minimal model obtained by neglecting chromatic-dispersion terms in 
the governing equations of the main text. This approximation reduces the governing equations to a pair of coupled transport equations describing the evolution of the pump and signal envelopes:

\begin{equation}
\begin{aligned}
    \frac{\partial a}{\partial z} + \beta^{(a)}_{1}\,\frac{\partial a}{\partial t}
    &=
     -\frac{\alpha^{(a)}}{2}a
    + \kappa\,a^{\ast} b,
    \\
    \frac{\partial b}{\partial z} + \beta^{(b)}_{1}\,\frac{\partial b}{\partial t}
    &=
        -\frac{\alpha^{(b)}}{2}b
    - \frac{1}{2} \kappa\,a^{2}.
\end{aligned}
\label{eq:transport_envelope}
\end{equation}

Eq.~\eqref{eq:transport_envelope} can be used to examine the nonlinear contribution to the walk-off. When the pulse trains evolve coherently as a whole, their effective
temporal motion can be characterized by the first moments (temporal centroids):
\begin{equation} \label{eq:RMS}
    T_{c}^{(a)}= \frac{\int{t |a(z,t)|^2 dt}}{\int{|a(z,t)|^2 dt}},\;\;\;T_{c}^{(b)} = \frac{\int{t |b(z,t)|^2 dt}}{\int{|b(z,t)|^2 dt}}.
\end{equation}
For an infinite periodic pulse train, a global first moment over the whole temporal axis is not defined. Eq.~\eqref{eq:RMS} should therefore be understood either over one temporal period or locally around a
representative pulse when the pulses are well separated.
Next, we compute how centroids ("centres of mass") evolve with propagation in each train:
\begin{equation} \label{eq:RMS2a}
    \frac{d T_{c}^{(a)}}{dz} = \beta^{(a)}_{1}  + \kappa \; \frac{\int{t (a^{*2} b + a^2 b^{*})dt}}{\int{|a(z,t)|^2 dt}} -\kappa \;   T_{c}^{(a)} \; \frac{\int{(a^{*2} b + a^2 b^{*})dt}}{\int{|a(z,t)|^2 dt}},
\end{equation}
\begin{equation} \label{eq:RMS2b}
    \frac{d T_{c}^{(b)}}{d z} = \beta^{(b)}_{1}  - \frac{\kappa}{2} \; \frac{\int{t (a^2 b^{*}+a^{*2} b)dt}}{\int{|b(z,t)|^2 dt}}+ \frac{\kappa}{2} \; T_{c}^{(b)} \; \frac{\int{(a^2 b^{*}+a^{*2} b)dt}}{\int{|b(z,t)|^2 dt}}.
\end{equation}
Let us define $R(t) = Re(a^2 b^*)$ we derive expression for the effective walk-off:
\begin{equation} \label{eq:U_eff}
    u_{c} = \beta^{(b)}_{1} - \beta^{(a)}_{1}  - 2 \kappa \; \frac{\int{(t-T_{c}^{(a)}) R(t)dt}}{\int{|a(z,t)|^2 dt}} - \kappa \; \frac{\int{(t -T_{c}^{(b)} ) R(t)dt}}{\int{|b(z,t)|^2 dt}}=  u_{lin}+ \Delta u_{NL}. 
\end{equation}
Eq.~\eqref{eq:U_eff} shows explicitly that asymmetric parametric energy exchange modifies the relative motion of the two pulse trains, giving an effective walk-off
\[
u_c=u_{\rm lin}+\Delta u_{\rm NL}.
\]
This nonlinear correction can bring the accumulated relative delay
toward the rational shift-register condition: 
\begin{equation}\label{eq:u_eff_rel}
 \frac{M}{N} = \frac{u_{c}}{\beta_1^{(a)}}.
\end{equation}

\section{Variational analysis for periodic pulse trains}
\label{sec:variational_periodic_train}
To get a qualitative understanding of the observed solutions, let us consider a lossless case (approximating systems with low cavity loss) with  $\alpha^{(a)}=\alpha^{(b)}=0$ and also neglect the third-order dispersion terms. 
In this case, basic equations  have a well-known Hamiltonian structure:
\begin{equation}
-i\frac{\partial a}{\partial z}
=
-\frac{\beta_2^{(a)}}{2}
\frac{\partial^2 a}{\partial t^2}
-i\kappa a^*b
=
\frac{\delta H}{\delta a^*},
\label{eq:var_a}
\end{equation}
and
\begin{equation}
-i\frac{\partial b}{\partial z}
=
iu\frac{\partial b}{\partial t}
-\frac{\beta_2^{(b)}}{2}
\frac{\partial^2 b}{\partial t^2}
+\frac{i\kappa}{2}a^2
=
\frac{\delta H}{\delta b^*}.
\label{eq:var_b}
\end{equation}
Here $\delta W/\delta s$ denotes the variational derivative of a functional $W$ with respect to $s$. 
The Hamiltonian $H=\int \widetilde{H}\,dt$ has the density:
\[ \tilde{H}[a,b] = \frac{\beta^{(a)}_{2}}{2} \left|\frac{\partial a}{\partial t}\right|^2 +\frac{\beta^{(b)}_{2}}{2} \left|\frac{\partial b}{\partial t} \right|^2+ \frac{i}{2} \;u\; (b^{*} \frac{\partial b}{\partial t} - b \frac{\partial b^{*}}{\partial t}) + \frac{i}{2} \kappa (a^2 b^{*} - a^{*2} b).  \]
Eqs.~\eqref{eq:var_a} -- \eqref{eq:var_b} have two well-known conserved quantities (integrals that do not change during evolution with $z$), Manley–Rowe invariant:
\[ N_{MR} = \int_{-\infty}^{\infty} {dt [ |a|^2 + 2 |b|^2]},\]
and a temporal momentum $P$:
\[P= - \frac{i}{2}\; \int_{-\infty}^{\infty} {dt \left( a^{*} \frac{\partial a}{\partial t}-
a \frac{\partial a^{*}}{\partial t}   + b^{*} \frac{\partial b}{\partial t}- b \frac{\partial b^{*}}{\partial t}   \right) }. \]
The corresponding Lagrangian density is
\[
{\cal L}
=
\frac{i}{2}\left(a^*a_z-aa_z^*\right)
+
\frac{i}{2}\left(b^*b_z-bb_z^*\right)
+ \tilde{H}[a,b]. \]

To obtain a reduced description consistent with the shift-register regime, we apply the variational approach to the periodic pulse trains. Note that this differs from the standard variational analysis for the isolated pulses (see e.g. \cite{Buryak2002, Karlsson1994, SKT1998} and references therein).

We use a variational approach to obtain a simplified description of numerically observed synchronized pulse trains. 
Let $T$ denote the common temporal period of the signal and pump pulse trains. The action is evaluated over one nonlinear section,
$0\le z\le L$, and over a temporal interval $t_0\le t\le t_1$. For a periodic solution one may choose a single temporal cell,
$t_1=t_0+T$. In the following, we additionally assume that the pulses
within each train are well separated. Consequently, linear contributions can be evaluated using one representative pulse, with the temporal integration extended to the whole real axis. Overlap with
neighbouring pulses of the same train is exponentially small and is
neglected to leading order.
 We keep, however, general integral limits from $t_0$ to $t_1$ below until this assumption is used. 
\begin{equation}
S[a,b] =
\int_0^L dz
\int_{t_0}^{t_1}{
{\cal L}\,dt}.
\label{eq:pt_action}
\end{equation}
Because the physical state is periodic in intensity, the value of the cell-integrated action is independent of the arbitrary origin $t_0$,
provided the complete Lagrangian density is periodic over the chosen cell. Admissible field variations satisfy either (approximately) the zero boundary conditions in the case of well-separated pulses or the corresponding quasi-periodic conditions. 
The temporal boundary terms generated by integration by parts vanish either because the representative pulses decay rapidly at the limits of integration or, for a finite temporal cell, because admissible
variations satisfy the corresponding periodic or quasi-periodic
boundary conditions. The optical intensities are periodic,  reflecting property of solutions obtained through numerical modeling
$|a(z,t+T)|^2=|a(z,t)|^2, \;\;
|b(z,t+T)|^2=|b(z,t)|^2$.

The Manley--Rowe quantity  is
${\cal N}_{\rm MR}
= E_a+2E_b$,
where
\begin{equation}
E_a
=
\int_{t_0}^{t_1}|a|^2dt,
\qquad
E_b
=
\int_{t_0}^{t_1}|b|^2dt.
\label{eq:pt_Eab}
\end{equation}
Thus $E_a$ and $E_b$ are the signal and pump pulse energies contained in one
temporal period of the train.
The temporal momentum per integration interval is
\begin{equation}
P
=
-\frac{i}{2}
\int_{t_0}^{t_1}
\left(
a^*a_t-aa_t^*
+
b^*b_t-bb_t^*
\right)dt.
\label{eq:pt_momentum}
\end{equation}

We now specify the trial functions. Let $T_s$ denote a characteristic
pulse width. Motivated by the numerically observed periodic states, we
represent the two fields as trains of well-separated pulses:
\begin{equation}
a(z,t)
=
A(z)
\sum_{n=-\infty}^{\infty}
F_a\!\left(
\frac{t-\tau_{a,n}(z)}{T_s}
\right)
\exp\left\{
i\phi_{a}(z)
+i\Omega_a(z)[t-\tau_{a,n}(z)]
\right\}.
\label{eq:pump_train}
\end{equation}
and
\begin{equation}
b(z,t)
=
B(z)
\sum_{n=-\infty}^{\infty}
F_b\!\left(
\frac{t-\tau_{b,n}(z)}{T_s}
\right)
\exp\left\{
i\phi_{b}(z)
+i\Omega_b(z)[t-\tau_{b,n}(z)]
\right\},
\label{eq:pump_train}
\end{equation}
where $F_a$ and $F_b$ describe the field-amplitude profiles of the
signal and pump pulses, respectively. For approximately uniform pulse trains,
$\tau_{a,n}(z)=\tau_{a}(z)+nT$, $\tau_{b,n}(z)=\tau_{b}(z)+nT$, $n=0,\pm 1, \pm 2,...$ 
where $T$ is the temporal period of the pulse train. Here $\tau_a(z)$ and $\tau_b(z)$ are translation coordinates of the two periodic solutions, respectively for signal and pump. For well-localized pulses their evolutions in $z$ coincide with the changes of
positions of signal and pump pulses  within a single temporal cell. 
As our focus here is on the relations between speed of the pulse train as a whole and changes of the phases, we consider $T_s$ to be constant in what follows. Generalisation to the case of pulse width varying with $z$ is straightforward, see \cite{Buryak2002, Karlsson1994, SKT1998}.
Although the reduction can be carried out for arbitrary localized profiles, we specialize from this point onward to equal-width unchirped Gaussian pulses,
\begin{equation}
F_a(x)=F_b(x)=\exp(-x^2/2).
\label{eq:Gaussian_ansatz}
\end{equation}
This choice allows all nonlinear overlap integrals to be evaluated analytically while retaining the essential dependence on relative delay, frequency mismatch, and phase.
The individual pulse energies are found as:
$E_a=\sqrt{\pi}\,A^2\;T_s$, and $E_b=\sqrt{\pi}\,B^2\;T_s$.
The Manley--Rowe invariant then is written as:
$N_{MR}=E_a+2E_b= \sqrt{\pi} \,(A^2+2\;B^2)\; T_s={\rm const_1}$.
The temporal moments per cell are $P_a=E_a\Omega_a$,
and $P_b=E_b\Omega_b$, so that $P=P_a+P_b=E_a \;\Omega_a+E_b\;\Omega_b= \sqrt{\pi}\,(A^2 \;\Omega_a + B^2\; \Omega_b)\; T_s={\rm const_2}$.
We can use in the variational approach either pair $A,B$ or $E_a,E_b$ as variables related to field power/amplitude. In what follows we will use pulse energies in the Lagrangian.
It is easy to see that the linear contributions to the reduced Lagrangian derived above can be evaluated using one representative pulse of each train. Since all pulses
within a given stationary train are related by temporal translation, the
energy, temporal momentum, dispersion, and walk-off contributions of the complete train are proportional to the number of pulses and reduce, after
normalization per pulse (or per temporal cell), to the expressions obtained in the section for a representative pulse. In the case of well-isolated pulses we will use $t_0= - \infty$ and $t_1 = \infty$. 

The nonlinear interaction requires a different treatment as signal and pump pulse trains are moving relative to each other. During propagation through the resonator a given signal pulse can overlap with several pump pulses as the two trains slip through
one another. Therefore, we can consider nonlinear interactions of a pump wave with a reference signal pulse.  

\subsection{Euler--Lagrange equations for the reduced variables}
We choose one signal pulse as a reference and label by $n=0$ the pump
pulse occupying the same temporal slot at the chosen reference plane.
The relative delay between the reference signal pulse and the $n$-th
pump pulse is then
\begin{equation}
\Delta_n(z)
=
\tau_{b,n}(z)-\tau_a(z)=\tau_{b}(z)-\tau_a(z)+nT=\Delta \,\tau + nT,
\label{eq:Delta_n}
\end{equation}
and its normalized form, for an equally spaced pump train,
\begin{equation}
\delta_n(z)
=
\frac{\Delta_n(z)}{T_s}= \frac{\Delta\tau(z)}{T_s}+n\,\frac{T}{T_s}=
\delta \tau(z)+n\,s_0,
\label{eq:delta_n_spacing}
\end{equation}

It is straightforward to derive
the reduced Lagrangian for the parameters of the trial functions:
\begin{align}
L_{\rm eff}
={}&
-E_a\phi_a'
-E_b\phi_b'
+
E_a\Omega_a\tau_a'
+
E_b\Omega_b\tau_b'
-uE_b\Omega_b
+
E_a\,\frac{\beta_2^{(a)}}{2}\, 
\left(
\frac{1}{2T_s^2}
+
\Omega_a^2
\right)
\nonumber\\
&+
E_b\,\frac{\beta_2^{(b)}}{2}\, 
\left( \frac{1}{2T_s^2}
+
\Omega_b^2
\right)
-
\Gamma\; E_a\, E_b^{1/2}\;
\sum_{n=-\infty}^{\infty}
I_{{\rm NL},n}(\phi_a,\phi_{b,n},\Omega_a,\Omega_{b,n},\tau_{b,n}-\tau_a).
\label{eq:Leff_multi}
\end{align}
Eq.~\eqref{eq:Leff_multi} is written in terms of the energies of the representative signal and pump pulses $E_a$ and $E_b$, so that the nonlinear coupling coefficient appearing in the reduced
Lagrangian is
\begin{equation}
\kappa A^2BT_s
= \Gamma E_a\sqrt{E_b}, \;\;\;
\Gamma
= \frac{\kappa}{\pi^{3/4}\;\sqrt{T_s}}.
\label{eq:Gamma_reduced}
\end{equation}
The nonlinear term $I_{{\rm NL},n}$ (for the $n$-th pump pulse) reads
\begin{equation}
I_{{\rm NL},n}
=
\int_{-\infty}^{\infty}
\exp(-x^2)
\exp\left[-\frac{(x-\delta_n)^2}{2}\right]
\sin(\Psi_n+Qx)\,dx = 
{\cal F}_n\sin\Theta_n,
\label{eq:gaussian_INL}
\end{equation}
where $Q=(2\Omega_a-\Omega_b)\,T_s$,
$\Psi_n = 2\phi_a-\phi_{b}
+\Omega_b\,(\Delta \tau +nT)=\Psi+n\;\Omega_b\,T$
and
\begin{equation}
\Theta_n
=
\Psi_n+\frac{Q\delta_n}{3}=2\phi_a-\phi_{b}+
\frac{2 }{3} (\Omega_a+\Omega_b) ( \tau_b - \tau_a) +n\,\frac{2}{3} (\Omega_a+\Omega_b) T =\Theta + n\; \Delta \Theta.
\nonumber
\end{equation}
Completing the square,
\begin{equation}
-x^2-\frac{(x-\delta_n)^2}{2}
=
-\frac{3}{2}
\left(x-\frac{\delta_n}{3}\right)^2
-\frac{\delta_n^2}{3},
\end{equation}
we can derive that
\begin{equation}
{\cal F}_n
=
\sqrt{\frac{2\pi}{3}}
\exp\left(
-\frac{\delta_n^2}{3}
-\frac{Q^2}{6}
\right)=\sqrt{\frac{2\pi}{3}}\,\exp\left(-\frac{Q^2}{6}-\frac{\Delta \tau^2}{3 T_s^2}\right)  \exp\left(
-  \frac{ 2n\,  T \Delta \tau}{3T_s^2} - \frac{n^2\,T^2}{3T_s^2} \right).
\label{eq:F_n}
\end{equation}
Thus, the nonlinear overlap integral  ${\cal I}=
\sum_n {\cal F}_n\sin\Theta_n={\cal I}(\tau_b-\tau_a,\Omega_a,\Omega_b,2\phi_a-\phi_{b}) $ can then be evaluated exactly:
\begin{equation}
{\cal I}=
\sqrt{\frac{2\pi}{3}}
\exp\left(-\frac{Q^2}{6}-\frac{\Delta \tau^2}{3 T_s^2}\right) \sum_n \exp\left(
-  \frac{ 2n\,  T \Delta \tau}{3T_s^2} - \frac{n^2\,T^2}{3T_s^2} \right) \sin{(\Theta + n \Delta \Theta)}.
\label{eq:I_gaussian_sum}
\end{equation}
\begin{figure}[t]
\centering
\includegraphics[width=0.70\linewidth]{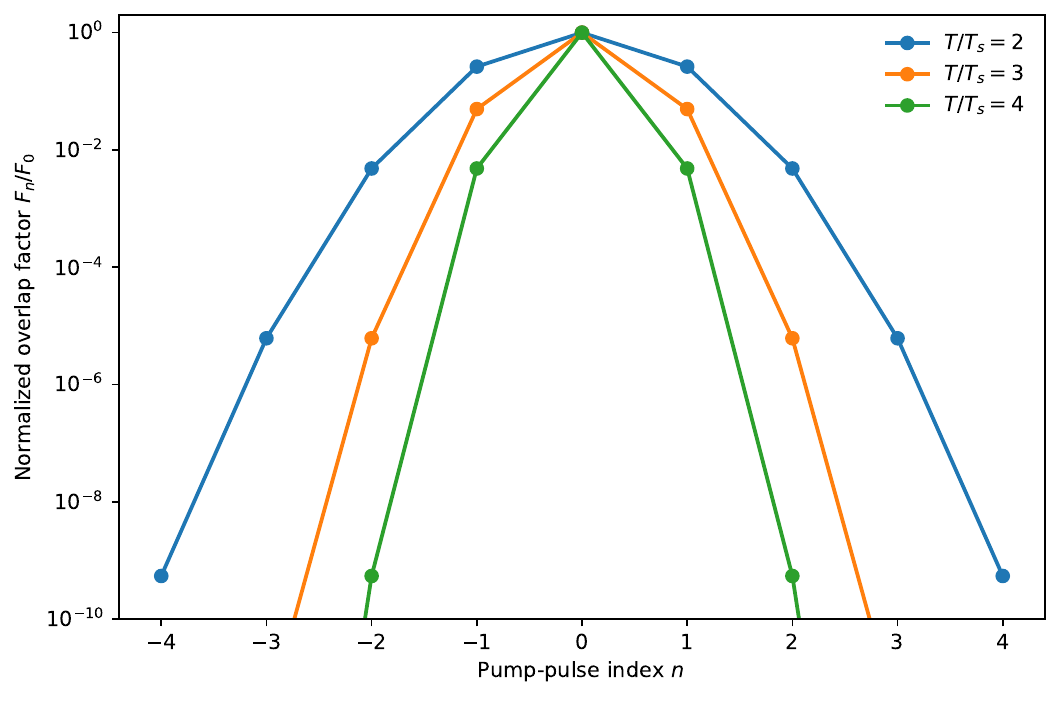}
\caption{
Normalized Gaussian overlap factor $F_n/F_0$ as a function of the pump-pulse
index $n$, calculated from Eq.~\eqref{eq:Fn_normalized} for
$T/T_s=2$, $3$, and $4$. The reference signal pulse is taken to be aligned
with the $n=0$ pump pulse, $\Delta\tau=0$. The rapid decay with $|n|$
demonstrates that, for well-separated pulses, only a small number of pump pulses nearest to the reference signal
pulse contribute appreciably to the nonlinear interaction at a given
propagation distance. 
}
\label{fig:Fn_overlap}
\end{figure}
The infinite sum in Eq.~\eqref{eq:I_gaussian_sum}, however, does not imply that all pump pulses contribute appreciably. Due to the rapid localization of pulse shapes, only pump pulses which overlap during propagation with the reference
signal pulse give a measurable nonlinear contribution. For well-localized pulses, the contribution of distant pump pulses is exponentially suppressed. 
For $\Delta\tau=0$, corresponding to temporal alignment of the reference
signal pulse with the $n=0$ pump pulse, the normalized overlap factor is
\begin{equation}
\frac{F_n}{F_0}
=
\exp\left[
-\frac{n^2}{3}
\left(\frac{T}{T_s}\right)^2
\right].
\label{eq:Fn_normalized}
\end{equation}
Thus, the number of pump pulses contributing appreciably to the nonlinear interaction is controlled primarily by the ratio of the pulse-train period to the pulse width, $T/T_s$. Fig.~\ref{fig:Fn_overlap} illustrates this localization for several representative values of $T/T_s$. As the pulse separation increases, the interaction rapidly becomes dominated by the
pump pulse nearest to the reference signal pulse.

Variation of the reduced Lagrangian with respect to the eight
collective coordinates
$\Omega_a,\Omega_b,\tau_a,\tau_b$,
$\phi_a,\phi_b, E_a, E_b$ gives the following closed set of evolution equations for the
parameters of the trial functions.  Equations for the energy evolution read:
\begin{equation}
\frac{dE_a}{dz}
=
2\Gamma \sqrt{\frac{2\pi}{3}} E_a\,E_b^{1/2}\, 
\exp\left(-\frac{Q^2}{6}-\frac{\Delta \tau^2}{3 T_s^2}\right) \sum_n \exp\left(
-  \frac{ 2n\,  T \Delta \tau}{3T_s^2} - \frac{n^2\,T^2}{3T_s^2} \right) \cos{(\Theta + n \Delta \Theta)}
\label{eq:Ea_gaussian}
\end{equation}
and
\begin{equation}
\frac{dE_b}{dz}
=
-\Gamma
\sqrt{\frac{2\pi}{3}} E_a\,E_b^{1/2}\, 
\exp\left(-\frac{Q^2}{6}-\frac{\Delta \tau^2}{3 T_s^2}\right) \sum_n \exp\left(
-  \frac{ 2n\,  T \Delta \tau}{3T_s^2} - \frac{n^2\,T^2}{3T_s^2} \right) \cos{(\Theta + n \Delta \Theta)}.
\label{eq:Eb_gaussian}
\end{equation}
Evidently, these equations conserve the Manley--Rowe invariant $E_a+2E_b=N_{MR}=const_1$.

After simple algebra, we get equations for $\tau_a$ and $\tau_b$: 
\begin{equation}
\frac{d\tau_a}{dz}
=
-\beta_2^{(a)}\Omega_a
+
\frac{2\Gamma \,E_b^{1/2}\, }{3}
 \sum_n {\cal F}_n \, \left[-Q T_s\,\sin{(\Theta_n)}+(\Delta \tau + n T)
\cos{(\Theta_n)}\right].
\label{eq:taua_gaussian}
\end{equation}

\begin{equation}
\frac{d\tau_b}{dz}
=
u-\beta_2^{(b)}\Omega_b
+
\frac{\Gamma E_a}{3E_b^{1/2}}
\sum_n {\cal F}_n \, \left[Q T_s\,\sin{(\Theta_n)}+2(\Delta \tau + n T)
\cos{(\Theta_n)}\right].
\label{eq:taub_gaussian}
\end{equation}
The frequency-momentum equations take the form:
\begin{equation}
\frac{d}{dz}(E_a\Omega_a)
=
\frac{2\Gamma E_a E_b^{1/2}}{3 T_s^2}
\sum_n{\cal F}_n
\left[
(\Omega_a+\Omega_b) \, T_s^2\cos\Theta_n
-
(\Delta \tau +n T)\sin\Theta_n
\right],
\label{eq:Omegaa_gaussian}
\end{equation}
and
\begin{equation}
\frac{d}{dz}(E_b\Omega_b)
=
-\frac{2\Gamma E_a E_b^{1/2}}{3 T_s^2}
\sum_n{\cal F}_n
\left[
(\Omega_a+\Omega_b) \, T_s^2\cos\Theta_n
-
(\Delta \tau +n T)\sin\Theta_n
\right],
\label{eq:Omegab_gaussian}
\end{equation}
Their sum obviously gives conservation of the temporal momentum
$E_a\Omega_a+E_b\Omega_b = P_{t}= const_2$.
Finally, variations with respect to the pulse energies $E_a$ and $E_b$
give the evolution equations for the corresponding phases:
\begin{equation}
\frac{d\phi_a}{dz}
=
\Omega_a\frac{d\tau_a}{dz}
+
\frac{\beta_2^{(a)}}{2}
\left(
\frac{1}{2 T_s^2}
+
\Omega_a^2
\right)
-
\Gamma E_b^{1/2}
\sum_n F_n\sin\Theta_n .
\label{eq:phi_a_gaussian}
\end{equation}
Similarly,
\begin{equation}
\frac{d\phi_b}{dz}
=
\Omega_b
\left(
\frac{d\tau_b}{dz}-u
\right)
+
\frac{\beta_2^{(b)}}{2}
\left(
\frac{1}{2T_s^2}
+
\Omega_b^2
\right)
-
\frac{\Gamma E_a }{2E_b^{1/2}}
\sum_n F_n\sin\Theta_n .
\label{eq:phi_b_gaussian}
\end{equation}

The above equations, therefore, form a closed reduced dynamical system for
$E_a,E_b,\Omega_a,\Omega_b,\tau_a,\tau_b,\phi_a$ and $\phi_b$.
The interaction of the reference signal pulse with the pump train enters through exponentially localized sums of the form
$\sum_n F_n(\delta_n,Q)(\cdots)$. At any fixed propagation distance, only terms with $|\delta_n|=O(1)$ contribute appreciably, while different pump pulses become active successively as the reference signal pulse slips through the train.

The two conserved quantities, the Manley--Rowe invariant and temporal momentum, can now be used to reduce the number of independent dynamical variables.

\subsection{Relative velocity and nonlinear walk-off}

The relative inverse velocity of the two pulse trains is
\begin{equation}
u_{\rm eff}
=
\frac{d}{dz}(\tau_b-\tau_a)=\frac{d \Delta \tau}{dz}.
\end{equation}
The effective walk-off reads
\begin{equation}
\boxed{
u_{\rm eff}
=
u+u_{dis}
+\Delta u_{\rm NL},
}
\label{eq:ueff_final}
\end{equation}
where walk-off modification due to dispersion  $u_{dis}=\beta_2^{(a)}\Omega_a
-\beta_2^{(b)}\Omega_b$, and nonlinear shift is:
\begin{equation}
\Delta u_{\rm NL}
=\frac{\Gamma E_a\sqrt{E_b}}{3}
\sum_n F_n
\Bigg[
2(\Delta\tau+nT)\cos\Theta_n
\left(
\frac{1}{E_b}-\frac{1}{E_a}
\right)
+
QT_s\sin\Theta_n
\left(
\frac{1}{E_b}+\frac{2}{E_a}
\right)
\Bigg].
\label{eq:DeltauNL_final}
\end{equation}
Eq.~\ref{eq:DeltauNL_final}  separates the effective walk-off correction into a dispersive contribution, $\Delta u_{\rm dis}=\beta_2^{(a)}\Omega_a-\beta_2^{(b)}\Omega_b$,
and a nonlinear contribution $\Delta u_{\rm NL}$. The latter is accumulated through successive interactions of the reference signal pulse with individual pump pulses.
 The contribution
of the $n$-th pump pulse is exponentially weighted by $F_n$ and contains
two terms: one associated with the relative temporal displacement of the
pulse centers and another arising from the phase variation across the
temporal overlap. Thus, for well-separated pulses, only a finite sequence
of pump pulses contributes appreciably during propagation through the
nonlinear section.

The nonlinear contribution to the accumulated relative temporal
displacement is
\begin{equation}
\Delta\tau_{\rm NL}
=
\int_0^L
\Delta u_{\rm NL}(z)\,dz.
\label{eq:DeltatauNL_final}
\end{equation}

A shift-register state is obtained when the total relative displacement
after propagation through the nonlinear section is commensurate with an
integer number of pulse periods,
\begin{equation}
\boxed{
\int_0^L u_{\rm eff}(z)\,dz
=MT.
}
\label{eq:SR_condition_final}
\end{equation}

Equivalently,
\begin{equation}
\int_0^L
\left[
u+\beta_2^{(a)}\Omega_a(z)
-\beta_2^{(b)}\Omega_b(z)
+\Delta u_{\rm NL}(z)
\right]dz
=MT.
\label{eq:SR_condition_explicit}
\end{equation}

In the quasi-stationary limit, where the frequency shifts and nonlinear velocity correction vary only weakly along the nonlinear section, Eq.~\eqref{eq:SR_condition_explicit} reduces to
\begin{equation}
u+\beta_2^{(a)}\Omega_a
-\beta_2^{(b)}\Omega_b
+\Delta u_{\rm NL}
\simeq
\frac{MT}{L}.
\label{eq:SR_condition_stationary}
\end{equation}
Eq.~\eqref{eq:SR_condition_stationary} provides a direct physical
interpretation of the shift-register state. 
This condition differs fundamentally from conventional velocity locking,
for which $u_{\rm eff}=0$. 
In the shift-register regime the interacting
waves need not propagate with identical effective velocities. The nonlinear interaction therefore does
not need to cancel the walk-off locally. Instead, it only needs to modify (renormalize) the
accumulated relative displacement so that it becomes commensurate with an
integer number $M$ of pulse periods.
Within this reduced description, a stationary shift-register state
corresponds to a relative periodic orbit satisfying (below $\Phi=2\phi_a-\phi_b$)
\begin{equation}
E_a(L)=E_a(0),\qquad
Q(L)=Q(0),\qquad
\Phi(L)=\Phi(0)\pmod{2\pi} ,
\end{equation}
while the relative temporal coordinate advances by an integer number of
pulse periods,
\begin{equation}
\Delta\tau(L)=\Delta\tau(0)+MT.
\end{equation}

The Gaussian nonlinear interaction depends on the frequency combination
$Q=T_s(2\Omega_a-\Omega_b)$, the relative temporal displacement
$\Delta\tau=\tau_b-\tau_a$, and the phase combination
$\Phi=2\phi_a-\phi_b$. 
The two conserved quantities,
$N_{\rm MR}=E_a+2E_b$,
and $P=E_a\Omega_a+E_b\Omega_b$,
allow all other variables  to be expressed
in terms of the four dynamical variables
$E_a,Q,\Delta\tau,\Phi$.
Using notation $q=Q/T_s$,
and the algebraic frequency combination
\begin{equation}
\Sigma
\equiv
\Omega_a+\Omega_b
=
\frac{
6P+(N_{MR}-3E_a)q
}{2N_{MR}},
\label{eq:Sigma_reduced}
\end{equation}
we can express the individual frequencies and the normalized pulse positions as follows:
\begin{equation}
\Omega_a=\frac{\Sigma+q}{3},
\qquad
\Omega_b=\frac{2\Sigma-q}{3},\;\;\delta_n=\frac{\Delta\tau+nT}{T_s}.
\label{eq:Omega_reconstruction}
\end{equation}
Then the phase entering the nonlinear overlap reads
\begin{equation}
\Theta_n
=\Phi
+\frac{2}{3}\Sigma\Delta\tau
+n\,
\frac{2}{3}\Sigma\, T.
\label{eq:Theta_reduced}
\end{equation}
For compactness, we introduce the overlap sums:
\begin{align}
C_0 &= \sum_n F_n\cos\Theta_n,
&
S_0 &= \sum_n F_n\sin\Theta_n,
\\
C_1 &= \sum_n \delta_nF_n\cos\Theta_n,
&
S_1 &= \sum_n \delta_nF_n\sin\Theta_n,
\end{align}
this makes 
$C_0,S_0,C_1,S_1$ functions only of
$(E_a,Q,\Delta\tau,\Phi)$.

The energy equation becomes
\begin{equation}
\frac{dE_a}{dz}
=
\Gamma E_a
\sqrt{2(N_{MR}-E_a)}\,C_0.
\label{eq:E_reduced_final}
\end{equation}
The frequency-mismatch equation is
\begin{equation}
\frac{dQ}{dz}
=
\frac{\Gamma}
{3\sqrt{(N_{MR}-E_a)/2}}
\left[
(3E_a-2N_{MR})Q\,C_0
-
2N_{MR} S_1
\right].
\label{eq:Q_reduced_final}
\end{equation}
The relative temporal coordinate obeys
\begin{equation}
\begin{aligned}
\frac{d\Delta\tau}{dz}
= u &+ \frac{Q}{3 T_s}
\left[
(\beta_2^{(a)}-2\beta_2^{(b)})
\, \frac{
6P T_s+(N_{MR}-3E_a)
}{2N_{MR}}
+
\beta_2^{(a)}+\beta_2^{(b)}
\right]\\
&+
\frac{\Gamma T_s}
{3\sqrt{(N_{MR}-E_a)/2}}
\left[
(3E_a-N_{MR})C_1
+
N_{MR}\,Q S_0
\right].
\end{aligned}
\label{eq:Deltatau_reduced_final}
\end{equation}

The evolution equation for the relative phase $\Phi=2\phi_a-\phi_b$
can also be expressed entirely in terms of the four variables used in the reduced equations:
\begin{equation}
\begin{aligned}
\frac{d\Phi}{dz}
&= D-\beta_2^{(a)}
\frac{(\Sigma+q)^2}{9}
+\frac{\beta_2^{(b)}}{2}
\frac{(2\Sigma-q)^2}{9}\\
&+\frac{2\Gamma T_s}
{9\sqrt{(N_{MR}-E_a)/2}}\times\left[(N_{MR}-3E_a)\Sigma+N_{MR}\,q \right]\times
C_1 \\
&+ \frac{\Gamma}
{\sqrt{(N_{MR}-E_a)/2}}\times
\left[
\frac{3E_a-2N_{MR}}{2}
-\frac{QT_s}{9}
\left\{
2N_{MR}\Sigma+(2N_{MR}-3E_a)\frac{Q}{T_s}
\right\}
\right] \times S_0.
\end{aligned}
\label{eq:Phi_reduced_final}
\end{equation}
Here $D=\beta_2^{(a)}/(2 \,T_s^2) -
\beta_2^{(b)}/(4 T_s^2)$.
These equations form a closed four-dimensional reduced dynamical system for four variables
$E_a,Q,\Delta\tau,\Phi$. 

Finally, the full numerical simulations show that, for the regimes considered here, the frequency mismatch $Q=T_s(2\Omega_a-\Omega_b)$ and the relative nonlinear phase
$\Phi=2\phi_a-\phi_b$ remain small. Moreover, the coefficient $D$ is
small for the parameters used in the simulations. This motivates the
additional approximation
$Q\simeq0,\;\;\Phi\simeq0,\;\;
\Omega_a\simeq\Omega_b\simeq0$.

Then $\Theta_n\simeq0$ and
\begin{equation}
F_n=
\sqrt{\frac{2\pi}{3}}
\exp\left[
-\frac{(\Delta\tau+nT)^2}{3T_s^2}
\right].
\end{equation}
The reduced dynamics collapse to two coupled equations,
\begin{equation}\label{eq:Ea_final_short}
\frac{dE_a}{dz}
=\Gamma E_a
\sqrt{2(N_{MR}-E_a)}
\sum_{n=-\infty}^{\infty}F_n ,
\end{equation}
and
\begin{equation}\label{eq:dtau_final_short}
\frac{d\Delta\tau}{dz}
=
u+
\frac{\Gamma (3E_a-N_{MR})}
{3\sqrt{(N_{MR}-E_a)/2}}
\sum_{n=-\infty}^{\infty}
(\Delta\tau+nT)F_n .
\end{equation}
Eq.~\eqref{eq:Ea_final_short} describes the parametric redistribution of energy between the two harmonics, whereas Eq.~\eqref{eq:dtau_final_short} shows explicitly how the same sequence of localized pulse encounters modifies the relative velocity of the two pulse trains. Despite the substantial reduction
from the full field equations to two coupled ordinary differential equations, this model retains the essential mechanism responsible for the nonlinear renormalization of the walk-off.
The conditions $Q\simeq0$ and $\Phi\simeq0$ should be regarded as an additional approximation appropriate to the numerically investigated states, rather than as an exact invariant reduction of the general
four-dimensional system.

\bibliography{References_Supplemental}